\documentclass[12pt]{article}

\usepackage[T1]{fontenc}                
\usepackage[USenglish]{babel}           
\usepackage[utf8]{inputenc}             
\usepackage[strict,autostyle]{csquotes} 
\usepackage{authblk}                    
\usepackage[font=small]{caption}        
\usepackage{graphicx}                   
\usepackage{microtype}                  
\usepackage{parskip}                    
\usepackage{lineno}                     
\usepackage{float}                      

\usepackage{subfig}
\usepackage{tabularx}
\usepackage{graphicx}
\usepackage[raggedrightboxes]{ragged2e}

\usepackage{geometry} 
\usepackage{hyperref} 
\usepackage{xcolor}   

\usepackage[backend=biber,sorting=none]{biblatex} 

\input{glyphtounicode}
\hypersetup{
    colorlinks  = true,
    urlcolor    = black!50!blue,
    linkcolor   = black!50!blue,
    citecolor   = black!50!blue,
    pdfauthor   = {David Oniani, Jordan Hilsman, Colonel (Ret.) Ronald K. Poropatich, MD, Jeremy C. Pamplin, MD, Yanshan Wang},
    pdftitle    = {From Military to Healthcare: Adopting and Expanding Ethical Principles for Generative Artificial Intelligence},
    pdfkeywords = {artificial intelligence, computer science, natural language processing, generative artificial intelligence},
    pdflang     = English,
    pdfpagemode = UseNone,
    pdfsubject  = {From Military to Healthcare: Adopting and Expanding Ethical Principles for Generative Artificial Intelligence}
}

\title{From Military to Healthcare: Adopting and Expanding Ethical Principles for Generative Artificial Intelligence}

\author[1]{David Oniani}
\author[1]{Jordan Hilsman}
\author[2]{Yifan Peng, PhD}
\author[3,4]{COL (Ret.) Ronald K. Poropatich, MD}
\author[5]{COL Jeremy C. Pamplin, MD}
\author[6,7]{LTC Gary L. Legault, MD}
\author[1,8,9,10,11]{Yanshan Wang, PhD\footnote{Corresponding author: yanshan.wang@pitt.edu}}

\affil[1]{\footnotesize Department of Health Information Management, University of Pittsburgh, Pittsburgh, PA}
\affil[2]{\footnotesize Department of Population Health Sciences, Weill Cornell Medicine, New York, NY}
\affil[3]{\footnotesize Division of Pulmonary, Allergy \& Critical Care Medicine, University of Pittsburgh, Pittsburgh, PA}
\affil[4]{\footnotesize Center for Military Medicine Research, University of Pittsburgh, Pittsburgh, PA}
\affil[5]{\footnotesize Telemedicine \& Advanced Technology Research Center, US Army, Fort Detrick, MD}
\affil[6]{\footnotesize Department of Surgery, Uniformed Services University, Bethesda, MD}
\affil[7]{\footnotesize Virtual Medical Center, Brooke Army Medical Center, San Antonio, TX}
\affil[8]{\footnotesize Intelligent Systems Program, University of Pittsburgh, Pittsburgh, PA}
\affil[9]{\footnotesize Department of Biomedical Informatics, University of Pittsburgh, Pittsburgh, PA}
\affil[10]{\footnotesize Clinical and Translational Science Institute, University of Pittsburgh, Pittsburgh, PA}
\affil[11]{\footnotesize University of Pittsburgh Medical Center Hillman Cancer Center, Pittsburgh, PA}

\date{}

\begin{document}
\maketitle


\vspace{-5ex}  
\begin{abstract}
\footnotesize
\noindent In 2020, the U.S. Department of Defense officially disclosed a set of ethical principles
to guide the use of Artificial Intelligence (AI) technologies on future battlefields. Despite stark
differences, there are core similarities between the military and medical service. Warriors on
battlefields often face life-altering circumstances that require quick decision-making. Medical
providers experience similar challenges in a rapidly changing healthcare environment, such as in the
emergency department or during surgery treating a life-threatening condition. Generative AI, an
emerging technology designed to efficiently generate valuable information, holds great promise. As
computing power becomes more accessible and the abundance of health data, such as electronic health
records, electrocardiograms, and medical images, increases, it is inevitable that healthcare will be
revolutionized by this technology. Recently, generative AI has captivated the research community,
leading to debates about its application in healthcare, mainly due to concerns about transparency
and related issues. Meanwhile, concerns about the potential exacerbation of health disparities due
to modeling biases have raised notable ethical concerns regarding the use of this technology in
healthcare. However, the ethical principles for generative AI in healthcare have been understudied,
and decision-makers often fail to consider the significance of generative AI. In this paper, we
propose GREAT PLEA ethical principles, encompassing governance, reliability, equity, accountability,
traceability, privacy, lawfulness, empathy, and autonomy, for generative AI in healthcare.
Furthermore, we introduce a framework for adopting and expanding practical ethical principles that
have been useful in the military to healthcare for generative AI, based on contrasting their ethical
concerns and risks. We aim to proactively address the ethical dilemmas and challenges posed by the
integration of generative AI in healthcare.
\\\\
\textbf{Keywords}: artificial intelligence, computer science, ethics, generative artificial intelligence, healthcare, military
\end{abstract}
\clearpage




\section{Introduction}

Artificial Intelligence (AI) plays an ever-increasing role in our daily lives and has influenced
fields from online advertising to sales and from the military to healthcare. With the ongoing AI
arms race in the Russia-Ukraine War, it is expected that AI-powered lethal weapon systems will
become commonplace in warfare~\cite{russiaukrainewar}. Although AI has shown promise in numerous
successful applications, there remains a pressing need to address ethical concerns associated with
these applications. There are dire consequences if an AI system selects an incorrect target
potentially killing non-combatants or friendly forces. Seeing the rapid emergence of AI and its
applications in the military, in 2020, the United States Department of Defense (DOD) disclosed
ethical principles for AI \cite{dod}. This document emphasized five core principles, aiming for
\textit{responsible}, \textit{equitable}, \textit{traceable}, \textit{reliable}, and
\textit{governable} AI \cite{dod}. To address the challenges posed by AI in the military, the North
Atlantic Treaty Organization (NATO) also released principles, including \textit{lawfulness},
\textit{responsibility} and \textit{accountability}, \textit{explainability} and
\textit{traceability}, \textit{reliability}, \textit{governability}, and \textit{bias mitigation}
\cite{nato}. Clearly, prominent military organizations demonstrate a cautious approach toward
adopting AI and are actively implementing measures to mitigate the risks associated with its
potential malicious uses and applications.

On the other hand, AI has had a direct impact on the healthcare industry, with discussions ranging
from the uses of AI as an assistant to medical personnel \cite{aidoccomparison, airadthrive,
aipathology} to AI replacing entire clinical departments \cite{airad, airadstars}. The use and
impact of AI in clinical Natural Language Processing (NLP) in the context of Electronic Health
Records (EHRs) have been profound \cite{aillmehr,aiehrophthalmology,aimedlitehr, singhal2023large}.
Similar to military organizations, the World Health Organization (WHO) has also released a document
discussing the ethics and governance of artificial intelligence for health \cite{who}.

Generative AI, as the name suggests, refers to AI techniques that can be used to create or produce
various types of new content, including text, images, audio, and videos. The rate of development of
generative AI has been staggering, with many industries and researchers finding its use in fields
such as finance \cite{genaifinance}, collaborative writing \cite{genaicollabwriting}, email
communication \cite{genaiemailcommunication}, and cyber threat intelligence \cite{genaicyber}.
Generative AI has also become an active area of research in the healthcare domain \cite{chatgptdiff,
gpt-3-medicallyaware}, such as report generation~\cite{Sun2023-jf} and evidence-based medicine
summarization~\cite{Peng2023-sm}.

Despite successful and potential AI applications, ethics has been one of the more controversial
subjects of discussion in the AI community, with diverging views and a plethora of opinions
\cite{gilbert2023bias, Birhane_2022}. Ethics deals with how one decides what is morally right or
wrong and is one of the pivotal aspects that we, as the AI research community, have to consider
carefully. Considering the recent emergence of generative AI models and their initial enthusiasm in
healthcare, our community must seriously consider ethical principles before integrating these
techniques into practical use. The military and healthcare are notably similar in many ways, such as
organizational structure, high levels of stress and risk, decision-making processes, reliance on
protocols, and dominion over life and death. Acknowledging these parallels and recognizing the
successful implementation of ethical principles in military applications, we propose to adopt and
expand these ethical principles to govern the application of generative AI in medicine.


\section{What is Generative Artificial Intelligence?}

Generative AI refers to AI that is used primarily for generating data, often in the form of audio,
text, and images. However, in this manuscript, we choose not to follow such a general definition and
instead, focus on a particular type of generative AI. In this section, we describe ``modern''
generative AI, discuss why it is important, and compare it to the term that has become so popular --
``AI.''

\begin{figure}[h]
    \centering
    \includegraphics[width=\linewidth]{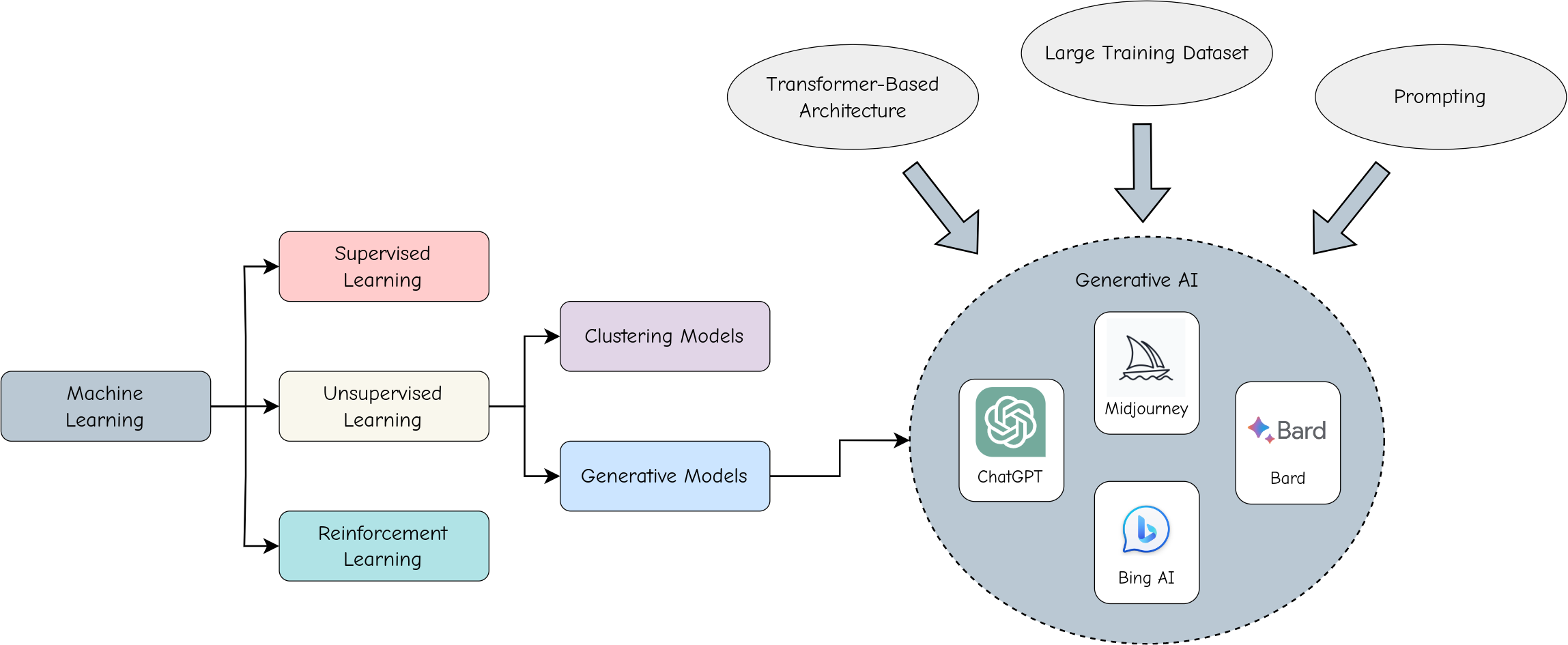}
    \caption{Machine Learning and Modern Generative Artificial Intelligence.}
    \label{aitree}
\end{figure}

Modern AI is dominated by Machine Learning (ML) methods, which leverage statistical algorithms and
large amounts of data to gradually improve model performance. ML methods could roughly be classified
into supervised, unsupervised, and reinforcement learning (Figure~\ref{aitree}). Supervised ML
relies on labeled input (supervision), while unsupervised learning needs no human supervision.
Reinforcement learning takes a different approach and, instead, attempts to design intelligent
agents by rewarding desired behaviors and punishing undesired ones. Popular generative AI models are
typically pre-trained in an unsupervised manner.

The pre-trained Generative AI models could generate novel and diverse outputs, including, but not
limited to, text, images, audio, or videos. Recently, the most popular generative AI model for
language generation is ChatGPT \cite{chatgpt}, which was reported to have an estimated 100 million
monthly active users in January 2023 \cite{chatgpt100mmactive}. The model architectures for ChatGPT,
previously known as GPT-3.5 \cite{chatgpt-gpt3.5}, and more recent GPT-4 \cite{gpt-4}, are built
upon the design principles of its GPT \cite{gpt} (Generative Pre-trained Transformer) predecessors,
GPT-2 \cite{gpt-2} and GPT-3 \cite{gpt-3}. Many state-of-the-art generative AI models, also known as
Large Language Models (LLMs), share a similar transformer-based architecture
\cite{attentionisallyouneed}.

The well-known generative AI models used for image generation, such as Stable Diffusion
\cite{stablediffusion} and DALL-E 2 \cite{dalle2}, employ a combination of the diffusion process
\cite{diffusionunifiedperspective} and a transformer-based architecture similar to the one used in
GPT models. All of the models are characterized by unsupervised training on very large datasets
\cite{zhao2023survey}. The same is true of models that generate images.

Most of these generative AI models also rely on a method called prompting \cite{liu2021pretrain},
which lets users input a natural language description of a task and uses it as a context to generate
useful information.

When referring to ``modern'' generative AI or simply generative AI, we are describing \textit{a
transformer-based machine learning model trained in an unsupervised manner on extensive datasets and
specifically optimized for generating valuable data through prompts}. This description also aligns
harmoniously with existing research and studies \cite{naturekather2022, genaicompletesurvey,
surveytexttoimage}.

While generative AI shows promising results, there are a number of problems with potential for
contributing to dangerous outcomes in healthcare, including:

\begin{itemize}
    \item Algorithmic bias \cite{chatgptbiaschallenges, chatgptpoliticalbias}
    \item Hallucination \cite{surveyhallucination, chatgpthallucination}
    \item Poor commonsense reasoning \cite{bian2023chatgpt, bang2023multitask}
    \item Lack of generally agreed model evaluation metrics \cite{chen2018metrics, thoppilan2022lamda}
\end{itemize}

All of these issues are common for generative AI in general, but more so in the healthcare domain,
where algorithmic bias may result in the mistreatment of patients \cite{gloria2022bias},
hallucination may carry misinformation \cite{peng2023study}, poor commonsense reasoning can result
in confusing interactions \cite{wei2023chainofthought}, and lack of general and domain-specific
metrics can make it difficult to verify the robustness of the system \cite{leiter2022explainable}.
Furthermore, in the context of healthcare, there are concerns about leaking Protected Health
Information (PHI) \cite{priyanshu2023chatbots} and machine empathy \cite{generativeaiempathy}.

Note that such concerns can also be present in other forms of AI, but given the practical
differences present in generative AI, the risks become elevated. First, due to the interactive
nature of generative AI, often paired with the ability to hold human-like dialogues (e.g., ChatGPT),
it can make misleading information sound convincing. Second, since generative AI models combine
various sources of large-scale data \cite{zhao2023survey}, the risk of training on biased data
sources increases. Third, the standard evaluation metrics, such as precision and recall, become
difficult to use and are less likely to reflect human judgment \cite{thoppilan2022lamda}. Finally,
due to its ease of use, generative AI has been widely adopted in many fields and domains, including
healthcare \cite{peng2023study}, which naturally increases the aforementioned risks.

Overall, the importance of ethical considerations for generative AI in healthcare cannot be
understated. From the human-centered perspective, the ultimate goal of generative AI is to enhance
and augment human's creativity, productivity, and problem-solving capabilities, which is well
aligned with the goal of healthcare in improving patient care. If the generative AI system is not
used ethically and does not reflect our values, its role as a tool for improving the lives of people
will greatly diminish.


\section{Applications in Military Versus Healthcare}

With the increasing prevalence of AI, it has been in the best interest of military organizations to
understand and integrate AI into their operations and strategies to be at the cutting edge of
security and technology in conflict or emergency. Various military AI technologies for generative
purposes have also been developed, including Intelligent Decision Support Systems (IDSSs) and Aided
Target Recognition (AiTC), which assist in decision-making, target recognition, and casualty care in
the field~\cite{donovan, atlas, doctrinaire}. Each of these uses of generative AI in military
operations reduces the mental load of operators in the field and helps them take action more
quickly. Just as military uses of AI can save lives on the battlefield, AI can help save lives by
assisting clinicians in diagnosing diseases and reducing risks to patient
safety~\cite{choudhury2020, bahl2017, dalal2019}. Uses of generative AI in healthcare help improve
the efficiency of professionals caring for patients. Applications of generative AI in healthcare
include medical chatbots, disease prediction, CT image reconstruction, and clinical decision support
tools~\cite{IntercomCB, siemenspredict, Willemink_CT, Bajgaine068373}. The benefits of such uses are
two-fold, in that they can help healthcare professionals deliver a higher level of care to their
patients, as well as improve the workload within clinics and hospitals.

People may question that developing AI models for military and healthcare purposes hinges on
distinct ideological underpinnings reflecting unique priorities. In the military context, AI models
are primarily designed to enhance the efficiency, precision, and strategic capabilities of both
defensive and offensive operations. The focus is on applications such as surveillance, target
recognition, cyber defense, autonomous weaponry, and battlefield analytics. Potential future uses of
AI for offensive actions such as coordinating drone attacks may oppose any healthcare principle, yet
is vital for the military strategy. The fundamental ideological perspective here is the protection
of national security interests, force multiplication, and minimizing human risk in conflict zones.

On the other hand, the use of AI in healthcare is driven by the principles of enhancing patient
care, improving health outcomes, and optimizing the efficiency of healthcare systems. The
development of AI models in this sector aims to personalize treatments, improve diagnostic accuracy,
predict disease progression, and streamline administrative tasks, among other uses. The central
ideology is the betterment of human health and well-being. While we acknowledge the different
ideological foundations in military and healthcare due to the contrasting objectives, we argue that
both military and healthcare sectors illustrate a compelling convergence of priorities for the
applications of generative AI.

Specifically, their shared focus on application validity, attention
to practical implementation, and prioritization of a human-centered approach have emerged as
significant commonalities. First, concerning application validity, both fields recognize the crucial
importance of robust, reliable generative AI systems. These systems need to function accurately and
rapidly under diverse, often challenging, conditions to fulfill their designated tasks, whether it
identifies potential security threats in a complex battlefield or detects subtle abnormalities in
medical images. Second, there is an evident emphasis on implementation. Beyond the theoretical
development of AI models, the critical question for both sectors centers around how these models can
be effectively incorporated into real-world systems, often involving multiple human and
technological stakeholders. Finally, a human-centered perspective is paramount. This means ensuring
that AI technologies augment, rather than replace, human decision-making capacities and are employed
in ways minimizing potential harm. In healthcare, this involves developing AI applications that can
improve patient outcomes and experience while supporting healthcare providers in their work. Thus,
these three factors represent key shared priorities in the utilization of generative AI across
military and healthcare contexts.

AI has been seamlessly woven into the military's technology fabric for several decades, serving as
the backbone for various advancements ranging from autonomous drone weapons to intelligent cruise
missiles \cite{david_lat_atlas, utegen2021}. The track record of robust results and reliable
outcomes in complex and high-risk environments implicitly engage with foundational ethical
principles. The ethical guidelines established from military AI implementations have provided a road
map for the incorporation of AI in healthcare scenarios. However, the integration of AI is
relatively new to the healthcare sector, and these ethical principles are not widely implemented and
are specifically designed for generative AI. While healthcare has begun to adopt generative AI
technologies more recently \cite{BOHR202025}, there are immense opportunities for this field to
glean ethical insights from the history of military application.

The U.S. military is shifting its focus from counterinsurgency operations in the Middle East to
Large Scale Combat Operations (LSCO) and an era of great power competition -- one where the United
States must be prepared to operate against peer and near-peer adversaries in high-end conflict
across air, land, sea, space, and cyber domains. Key to the modernization of the U.S. military is
the development of Robotic Autonomous Systems (RAS) built on AI-focused developed capability
\cite{largescalecombat}. As autonomy and AI capabilities improve, the vision is that robotics could
move from being tools to being teammates. Using AI to quickly analyze massive amounts of data could
shorten the decision-making process. However, soldiers would have to trust the identifications made
by these algorithms, which require well-developed human systems integration \cite{robotsintohumans}.

The Safety of Human-AI-Machine teams means agents trust each other, can understand and explain their
reasoning to each other, and further co-evolve to be better teammates as they work together.
Successful human teams rely on humans’ ability to learn about each other, predict one another's
intents, and communicate, more and more succinctly as they co-evolve. The U.S. military expects that
Human-AI-Machine teams will have amplified capability over human-only teams, provide new kinds of
capabilities, and, most importantly, they must be ``high-confidence.'' That is, they must have
provable, or predictable, attributes of performance, must be tolerant to faults in the environment
and configuration, and be able to resist kinetic and cyber attacks \cite{robotsintohumans}. The U.S.
military focus on AI-enhanced RAS for warfighter needs at the forward edge of the battlefield. The
inherent need to ensure the incorporation of strong ethical standards is also key to new development
of AI-enhanced military medical care in a pre-hospital battlefield setting.


\section{Identifying Ethical Concerns and Risks}

A RAND corporation study raised various concerns about the use of AI in warfare, shown in Figure 2.3
of the research report \cite{RR-3139-1-AF}. These concerns fall into the following categories:
increasing risk of war, increased errors, and misplaced faith in AI. Although AI can allow personnel
to make decisions and strategies more quickly, some experts consider this a downside, as actions
taken without proper consideration could have serious repercussions, like increasing the risk of war
\cite{RR-3139-1-AF}. International standards for warfare like the Law of Armed Conflict (LOAC) and
Geneva Conventions lay out guidelines for target identification specifying that attacks must first
distinguish between combatant and noncombatant targets before taking action to minimize harm to
civilians \cite{loac, ihl}. Because combatants are not always identifiable visually, some claim that
reading body language to differentiate a civilian from a combatant necessitates a Human-In-The-Loop
(HITL) decision-making process \cite{docherty_2023}.

Maintaining data privacy for users of generative AI technologies is critical, as both patient data
and military data are highly sensitive, and would be damaging if leaked \cite{genaiaprivacy}. If an
AI implementation collects PHI, it should be secure against breaches, and any disclosures of this
protected data must comply with Health Insurance Portability and Accountability Act (HIPAA)
guidelines \cite{HIPAAJournal_2023}. These implementations must experience few errors as healthcare
is a safety-critical domain where the patient harm is unacceptable \cite{patel2015cognitive}, and
errors in these systems or algorithms could cause more harm than any physician would be capable of,
as many hospitals and clinics would be using the same systems and experiencing the same
errors~\cite{brookings2022}.

\begin{figure}[h]
    \centering
    \includegraphics[width=0.45\linewidth]{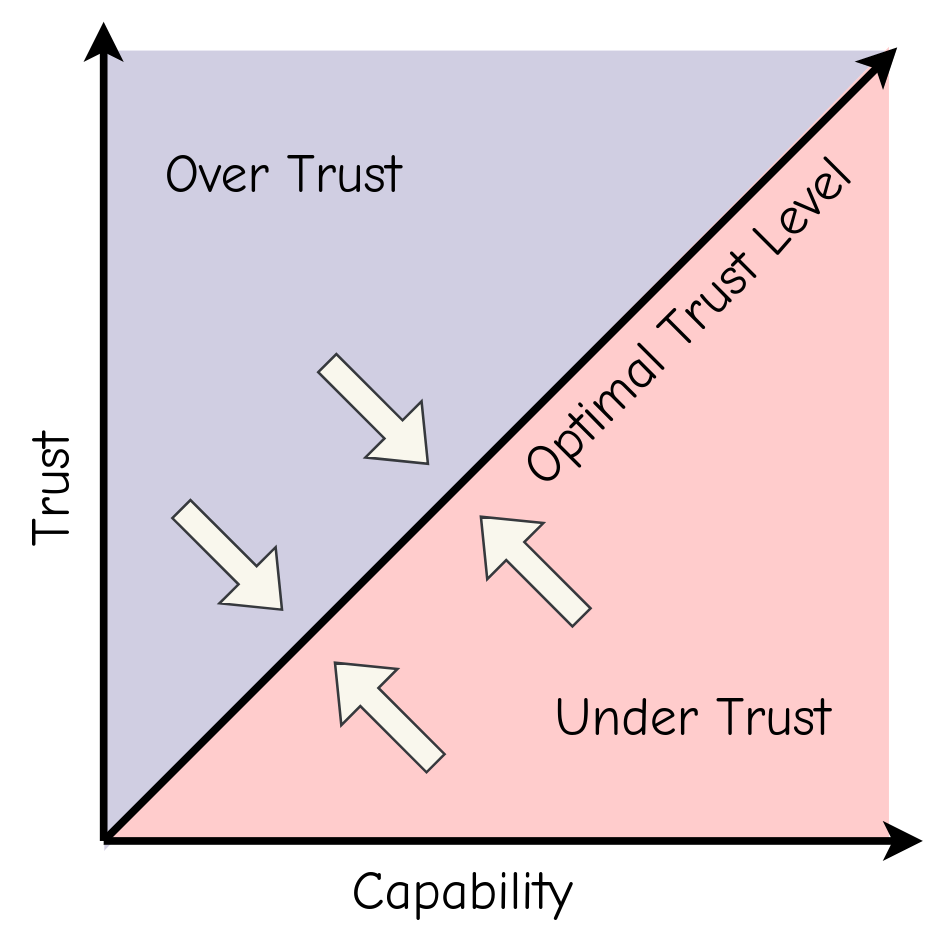}
    \caption{Optimization of Trust in AI}
    \label{trustgraph}
\end{figure}

Additional concerns present in the military and healthcare are trust between humans and AI and the
lack of accountability. When there exists human-and-AI collaboration to perform a task, trust must
be optimal, as shown in Figure \ref{trustgraph}. Too much trust in AI systems can lead to overuse of
the AI when it is not in the best interest of patients or operators \cite{lyons2012}, and too little
trust can lead to under-use of the system when it would be better to use it \cite{asan2019}. In both
situations, the root cause is operators not knowing the capabilities and limitations of the systems
they interact with \cite{Lewis2018}. Misuse can lead to non-typical errors, such as fratricide in
the military or patient harm in a hospital \cite{Hawley2007LookingBA, parikh2019}. While the AI must
be transparent in its decision-making, the use of AI must be accompanied by sufficient education on
the use and limitations of AI systems so that operators are less likely to make dangerous errors. A
lack of accountability can possibly arise in military or healthcare use of generative AI because
military operators or clinicians do not have direct control over the actions determined by the AI.

In a research report by Rand Corporation \cite{RR-3139-1-AF}, authors showed (Figures 7.8, 7.9 in
the report) that the general public views autonomous systems taking military action with human
authorization favorably while strongly disagreeing with combat action without human authorization.
The parallel can be drawn with healthcare, where patients express concerns over the use of AI for
medical purposes without human (e.g., physician, nurse, etc.) involvement~\cite{Richardson2021}.
These results could be due to the perceived lack of accountability, which is considered something
that could entirely negate the value of AI, as a fully autonomous system that makes its own
decisions distances military operators or clinicians from the responsibility of the system's actions
\cite{russell2015}. In healthcare, it is critical that the systems are transparent due to their
proximity to human lives and that patients understand how clinicians use these recommendations. The
burden of accountability in the healthcare sector falls to both the clinicians and the developers of
the AI systems, as the decisions made are a product of the algorithm, and the use of these
recommendations falls to the clinicians \cite{habli2020}.

Finally, ethical concerns of equity, autonomy, and privacy regarding the use of generative AI must
also be considered. In healthcare settings, biased algorithms or biased practices can lead to
certain patient groups receiving lower levels of care \cite{obermeyer2019}. Biased outcomes could be
due to biased algorithms, poor data collection, or a lack of diversity \cite{obrien2022}. There must
be minimal bias in developing AI systems in healthcare, both in the algorithm and the data used for
training. Furthermore, if known, the sources of bias must also be disclosed to ensure transparency
and prevent inappropriate use. The issue of human autonomy when developing generative AI is
especially pertinent in healthcare, as both patient and clinician autonomy must be respected
\cite{char2018}. It is crucial that a framework is accepted to prevent any data breaches and ensure
security measures are up to date and robust.

These risks and ethical concerns surrounding generative AI in military and healthcare applications
necessitate principles for the ethical use of AI. One of the earliest sets of principles published
for responsible development and use of AI comes from Google, who did so in response to their
employees petitioning their CEO as they disagreed with Google working with the DOD on Project Maven
to assist in identifying objects in drone images~\cite{frisk_2018,shane_wakabayashi_2018} in 2018.
These principles outline how Google will develop AI responsibly and state what technologies they
will not create, like those that cause harm or injure people, provide surveillance that violates
international policies, and any technologies that go against international law and human rights
\cite{google_ai}. By examining the differences and similarities between risks and ethical concerns
in military and healthcare applications of generative AI, we can establish guiding principles for
the responsible development and use of generative AI in healthcare.


\section{GREAT PLEA Ethical Principles for Generative AI in Healthcare}

As AI usage has spread throughout the military and other fields, many organizations have recognized
the necessity of articulating their ethical principles and outlining the responsibilities associated
with applying AI to their operations. There are numerous principles of AI ethics published by
various organizations, such as the U.S. Department of Defense \cite{dod}, NATO \cite{nato}, the
American Medical Association (AMA) \cite{amaprinciples}, the World Health Organization (WHO)
\cite{who}, and the Coalition for Health AI (CHAI) \cite{chai}. The AI ethical principles for DOD
and NATO are similar, with NATO having an added focus on adherence to international law. For the
development of AI for healthcare, the WHO has published its own ethical principles, including
protecting human autonomy, human well-being and safety, transparency and explainability,
responsibility and accountability, inclusiveness, and responsive development. The AMA policy is much
the same, promoting the development that is user-centered, transparent, reproducible, avoids
exacerbating healthcare disparities, and safeguards the privacy interests of patients and other
individuals. Finally, there is the Blueprint for an AI Bill of Rights published by the U.S. Office
of Science and Technology Policy (OSTP) \cite{thewhitehouse_2023}, which has provisions for AI
systems to be safe and effective, protected against algorithmic discrimination, protect user data,
have accessible documentation, and offer human alternatives.

Among the various sets of principles, we see common themes such as accountability and human
presence. The DOD and NATO both emphasize the importance of integrating human responsibility into
the development and life cycle of an AI system, as well as ensuring these systems are governable to
address errors that may arise during use. The AMA and WHO policies both highlight a human-centered
design philosophy protecting human autonomy and explicitly mention the need for inclusiveness and
equity in the healthcare use of AI to prevent care disparity. These principles each provide unique
perspectives for developing AI for healthcare use. However, no set of principles encompasses all
ethical concerns that healthcare providers or patients may have \cite{naik2022}. Adopting the
principles of the DOD and NATO is advantageous due to each principle's practical definition. These
principles are outlined with a focus on what actions can be taken by personnel developing AI
systems, and how end-users would interact with the systems.

The existing principles establish a good foundation for the ethical development and utilization of
AI in healthcare. However, action must be taken to tailor these principles for generative AI. By
examining the risks and concerns surrounding the use of generative AI in healthcare, comparing them
to the risks and concerns of generative AI in the military, and by expanding these principles, we
can have a set of principles that fulfill our needs \cite{Pifer_2023}.

\begin{figure}[H]
    \centering
    \includegraphics[width=0.64\linewidth]{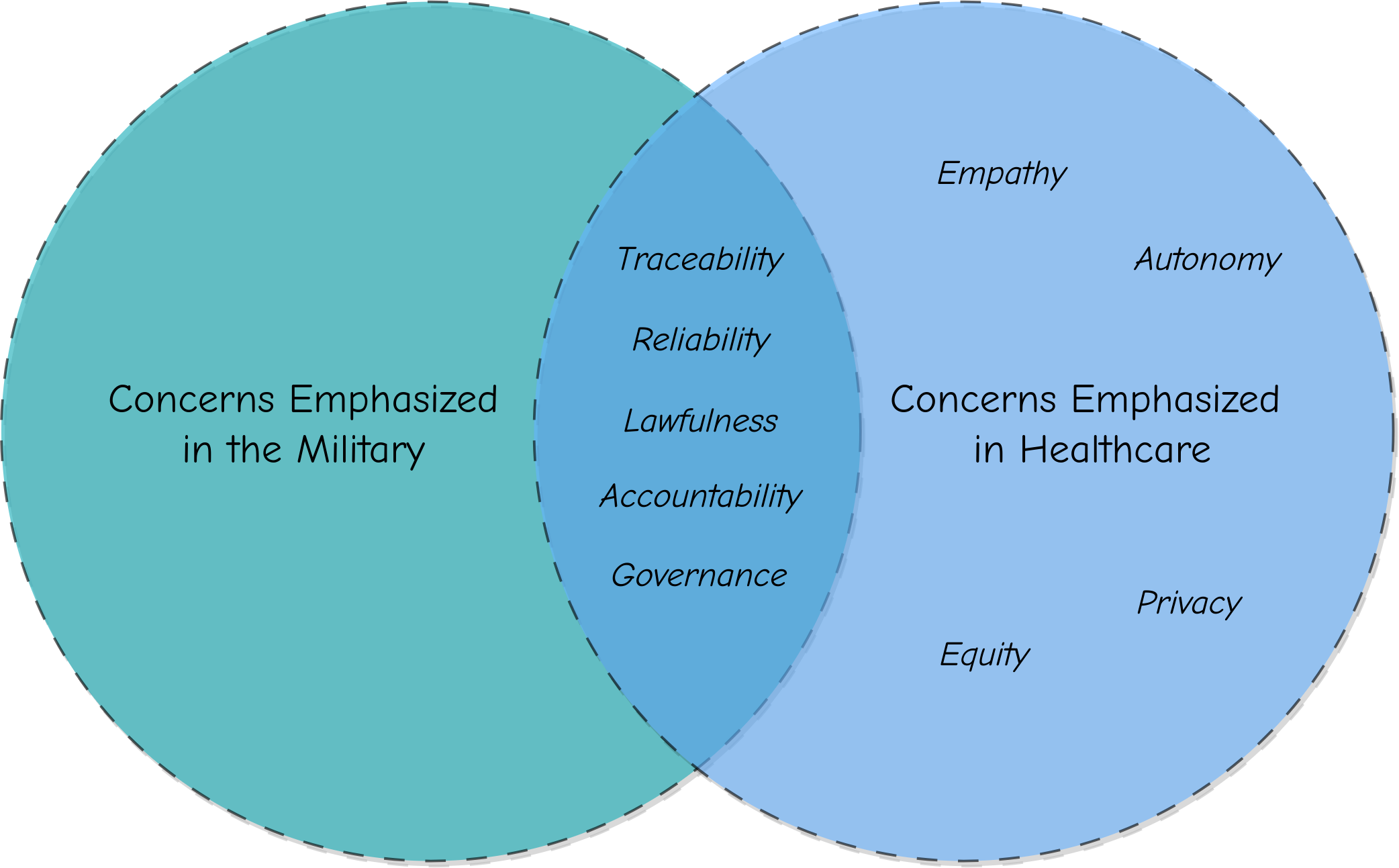}
    \caption{Adoption and Expansion of Existing Ethical Principles from Military to Healthcare}
    \label{principlesvenn}
\end{figure}

\begin{figure}[H]
    \centering
    \includegraphics[width=\linewidth]{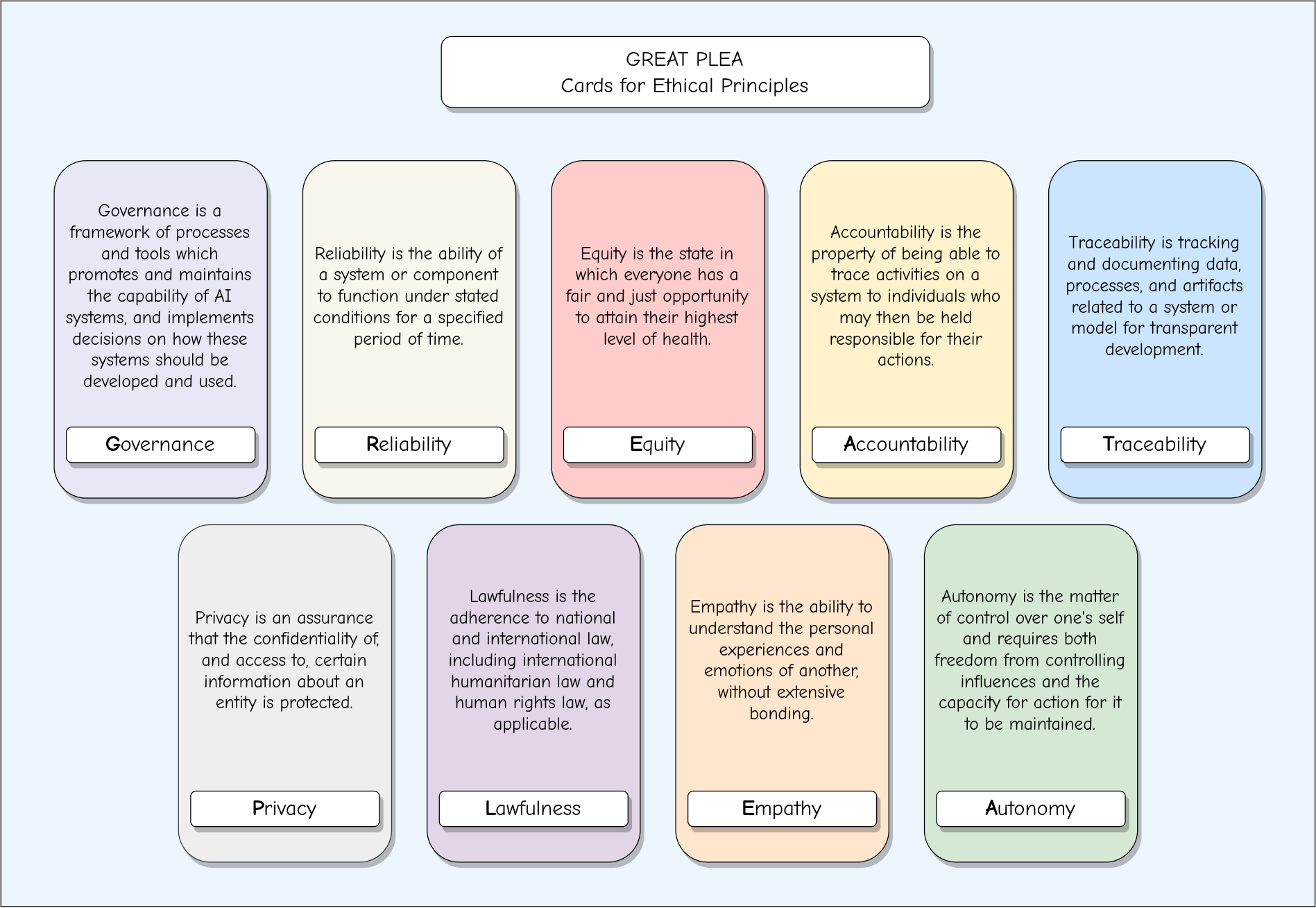}
    \caption{GREAT PLEA Cards for Ethical Principles}
    \label{great_plea_cards}
\end{figure}

Figure \ref{principlesvenn} shows the framework that we used for adopting and expanding ethical
principles, established by various organizations, for the healthcare applications of AI. Where
similarities are present in the concerns between military and healthcare use of generative AI, it is
possible to adopt principles for use, such as traceability, reliability, lawfulness, accountability,
and governance. In instances when healthcare has unique circumstances or requires additional nuance,
the principles related to those matters must be expanded to fit into the world of medicine, such as
empathy, equity, autonomy, and privacy. In summary, we propose the GREAT PLEA ethical principles:
\underline{G}overnance, \underline{R}eliability, \underline{E}quity, \underline{A}ccountability,
\underline{T}raceability, \underline{P}rivacy, \underline{L}awfulness, \underline{E}mpathy, and
\underline{A}utonomy, which demonstrate our great plea for the community to prioritize these ethical
principles when implementing and utilizing generative AI in practical healthcare settings. Figure
\ref{great_plea_cards} shows the summary cards for the GREAT PLEA ethical principles. In the
following, we will delve into a comprehensive explanation of each individual principle.

\subsection{\textit{Governance}}

Governance is a framework of processes and tools which promotes and maintains the capability of AI
systems, and implements decisions on how these systems should be developed and used
\cite{humanaigovernance}. Standards for the governance of AI systems, as established by the DOD and
NATO, emphasize the importance of ensuring that while AI systems fulfill their intended functions,
humans must retain the ability to identify and prevent unintended consequences. In the event of any
unintended behavior, human intervention to disengage or deactivate the deployed AI system should be
possible. These standards can be adopted for the use of generative AI in healthcare. Due to the
potential of widespread implementation of generative AI systems, where numerous hospitals may be
using the same systems, these standards must be considered~\cite{brookings2022}. Suppose a
generative AI system, deployed across multiple clinics, poses a risk of harm to a patient. In that
case, it is crucial to recognize that numerous patients across clinics could be vulnerable to the
same error. Risk to patients amplifies as healthcare expands to patient homes with remote patient
monitoring or with online tools outside the clinic. Ideally, humans, whether they develop or
implement the system, should possess the capability to deactivate it without disrupting the regular
patient care activities in the clinics. There must be explicit guidelines for monitoring generative
AI systems for potential errors, deactivation to prevent more damage when an error occurs, remedying
errors, and interaction to reduce operator errors. With these guidelines in place, personnel in
charge of the system can quickly be notified of any unintended behavior and respond quickly and
appropriately.

\subsection{\textit{Reliability}}

Reliability is the ability of a system or component to function under stated conditions for a
specified period of time~\cite{nist_reliability}. The proximity of generative AI to patient
well-being necessitates standards for reliability to minimize potential errors that could lead to
accidents \cite{surveyhallucination}. The generative AI models should have explicit and well-defined
clinical use cases. A generative AI model designed for disease prediction needs to have a clear
definition of the use situation and patient criteria. In addition, such generative AI models should
be safe, secure, and effective throughout their life cycles. Generative AI models should be
demonstrated to be at least as safe as human decision making alone and not cause undue harm.
Existing generative AI models suffer from hallucination and output variations, undermining their
ability to produce reliable outputs. These shortcomings can adversely affect the trust between
physicians and generative AI systems. Adopting the DOD's principle for reliability can establish use
cases for AI applications and monitor them during development and deployment to fix system failures
and deterioration. Having a thorough evaluation and testing protocol against specific use cases will
ensure the development of resilient and robust AI systems, and help minimize system failures as well
as the time needed to respond to these errors.

\subsection{\textit{Equity}}

Equity is the state in which everyone has a fair and just opportunity to attain their highest
level of health~\cite{cdc_equity}. Due to the importance of health equity and the ramifications of
algorithmic bias in healthcare, we call for adjustments to this principle. There already exists
inequity in healthcare. The generative AI models should not exacerbate this inequity for
marginalized, under-represented, socioeconomically disadvantaged, low education, or low health
literacy groups~\cite{Aquinojme2022108850}, but rather incorporate their unique social situations
into future AI models to insure equity. Generative AI must be developed with efforts to mitigate
bias by accounting for existing health disparities. Without this consideration, generative AI
systems could erroneously recommend treatments for different patients \cite{hoffman2016}. Expansion
of the principle for equity must set standards for evaluation metrics of algorithmic fairness so
that deployed AI systems will not reinforce healthcare disparity.

\subsection{\textit{Accountability}}

Accountability is the property of being able to trace activities on a system to individuals who may
then be held responsible for their actions~\cite{accountability_nist}. To ensure accountability and
human involvement with AI in healthcare, the principle of Responsibility and Accountability outlined
by NATO \cite{nato} states that they will develop AI applications mindfully and integrate human
responsibility to establish human accountability for actions taken by or with the application. A
study of patient attitudes toward AI showed the importance of accountability in gaining patient
trust when using AI in healthcare \cite{Robertson_Woods_Bergstrand_Findley_Balser_Slepian}. This
assurance of accountability is crucial when a clinician is using generative AI to help treat a
patient, as without proper measures for human accountability, the patient may feel the clinician is
not invested in the car they are delivering \cite{habli2020accountability}. We can adopt this
principle for the ethical use of generative AI in healthcare, and ensure that human involvement is
maintained when more powerful generative AI systems such as ChatGPT or clinical decision support
systems are in use.

\subsection{\textit{Traceability}}

Traceability is tracking and documenting data, processes, and artifacts related to a system or model
for transparent development~\cite{bdcc5020020}. Addressing the issue of optimizing trust between
healthcare professionals and the AI they interact with can be done by adopting the principle of
Traceability. This way, the personnel working with AI will understand its capabilities,
developmental process, methodologies, data sources, and documentation. Furthermore, providing
personnel with the understanding of an AI system capabilities and the processes behind its actions,
will also improve system reproducibility, allowing for seamless deployment across healthcare
systems. This is important for generative AI systems in healthcare because of their nature of being
a black box system. This high-level understanding will help optimize trust, as operators will be
aware of the capabilities and limitations of the AI systems they work with and know the appropriate
settings for use \cite{li2022trustworthy}. With generative AI becoming more prevalent in healthcare,
proper documentation is required to ensure all end users are properly educated on the capabilities
and limitations of the systems they interact with. The generation process of generative AI models
should be transparent. The references or facts should be provided together with answers and
suggestions for clinicians and patients. Data sources used to train these models and the design
procedures of these models should be transparent too. Furthermore, the implementation, deployment,
and operation of these models need to be auditable, under the control of stakeholders in the
healthcare setting.

\subsection{\textit{Privacy}}

Privacy is an assurance that the confidentiality of, and access to, certain information about an
entity is protected~\cite{privacy_nist}. Privacy is necessary in most military and medical
applications of healthcare due to their confidential nature. Generative AI systems in healthcare
must be HIPAA compliant for data disclosures, and secure to prevent breaches and developers should
be advised how healthcare data should train systems for deployment. HIPAA compliance requires a
regular risk assessment to determine how vulnerable patient data is \cite{hipaariskanalysis}, thus a
clinic utilizing generative AI systems in healthcare would have to determine if these systems are
weak points in their technology network. For example, the utilization of generative AI models
presents potential privacy breach risks, including prompt injection~\cite{perez2022ignore}, where
malicious actions could be conducted by overriding an original prompt, and
jailbreak~\cite{liu2023jailbreaking}, where training data could be divulged by eliciting generated
content. Furthermore, the capabilities of generative AI to process personal data and generate
sensitive information make it crucial for these systems to be secure against data breaches and
cyber-attacks. Ensuring these systems are developed with data privacy and security in mind will
assist in keeping protected patient information secure. Having these robust measures in place to
maintain the privacy of the sensitive data collected and made by AI systems is crucial for the
well-being of patients and for building trust with patients.

\subsection{\textit{Lawfulness}}

Lawfulness is the adherence to national and international law, including international humanitarian
law and human rights law, as applicable~\cite{lawfulness_nato}. This can be adopted for the use of
generative AI in healthcare. The laws that must be adhered to are not laws of conflict, but rather
those related to healthcare. Different states in the U.S. may establish different laws for AI
systems that must be heeded for deployment in those areas \cite{Team_2022}. Generative AI systems in
healthcare also face legal challenges surrounding safety and effectiveness, liability, data privacy,
cybersecurity, and intellectual property law \cite{naik2022}. A legal foundation must be established
for the liability of action taken and recommended by these systems, as well as considerations for
how they interact with cybersecurity and data privacy requirements of healthcare providers.
Generative AI for healthcare must be developed with these legal challenges in mind to protect
patients, clinicians, and AI developers from any unintended consequences.

\subsection{\textit{Empathy}}

Empathy is the ability to understand the personal experiences and emotions of another, without
extensive bonding~\cite{moudatsou2020}. A principle for empathy is not directly referenced in any
guidelines by the DOD or NATO. However, by emphasizing the need for human involvement in the
treatment of patients, it is possible to create a framework for human involvement in generative AI
applications to prevent gaps in accountability and ensure patients receive care that is empathetic
and helpful \cite{generativeaiempathy}. Chatbots like ChatGPT have been shown to lack empathy,
further reinforcing the need for a principle defining empathy for generative AI in healthcare
\cite{asch2023}. An empathetic relationship between provider and patient brings several benefits to
both the patient and the clinic treating them, such as better patient outcomes, fewer disputes with
healthcare providers, higher patient satisfaction, and higher reimbursement \cite{moudatsou2020}.

\subsection{\textit{Autonomy}}

Autonomy is the matter of control over one's self and requires both freedom from controlling
influences and the capacity for action for it to be maintained~\cite{Holm332, AMA_2016}. The
protection of autonomy needs to be ensured when using generative AI in healthcare. The more powerful
generative AI systems become, the more concern arises that humans do not control healthcare systems
and care decisions \cite{attentionisallyouneed}. Protecting human autonomy means that patients
receive care according to their preferences and values and that clinicians can deliver treatment in
the manner they want, without being encroached upon by the generative AI system. If autonomy in
decision-making is not patient-focused, the potential for adverse events and poor clinical outcomes
will surely follow \cite{applin2016}. By including provisions for protecting autonomy in using
generative AI in healthcare, doctor-patient relations improve, and care quality is ultimately
improved \cite{docpatrelations}.

We propose the GREAT PLEA ethical principles in the hope of addressing the ethical concerns of
generative AI in healthcare, as well as the distinction between generative AI and ``general'' AI.
This will be achieved by addressing the elevated risks mentioned previously in the paper. Generative
AI necessitates guidelines that account for the risk of misinformation, ramifications of bias, and
difficulty of using general evaluation metrics. Considering the widespread nature of generative AI
and its risks, these ethical principles can protect patients and clinicians from unforeseen
consequences. Following these principles, generative AI can be continuously evaluated for errors,
bias, and other concerns that patients or caregivers may have about their relationship with AI in
their field. These principles can be enforced through cooperation with lawmakers and the
establishment of standards for developers and users, as well as a partnership with recognized
governing bodies within the healthcare sector, such as the WHO or AMA.

\section{Conclusion}

Generative AI has great potential to enhance and make high-quality healthcare more accessible to
all, leading to a fundamental transformation in its delivery. Challenges posed by AI in healthcare
often mirror those encountered in military. In this paper, we proposed GREAT PLEA ethical
principles, encompassing nine principles designed to promote the ethical deployment of generative AI in healthcare. 
The present moment urges us to embrace these principles, foster a closer collaboration between humans and technology, and effect a radical enhancement in our healthcare system.

\section{Acknowledgements}

The authors would like to acknowledge support from the University of Pittsburgh Momentum Funds,
Clinical and Translational Science Institute Exploring Existing Data Resources Pilot Awards, the
School of Health and Rehabilitation Sciences Dean’s Research and Development Award, the National
Institutes of Health through Grant UL1TR001857, 4R00LM013001, R01LM014306, and the National Science Foundation
through Grant 2145640. The sponsors had no role in study design; in the collection, analysis, and interpretation of data; in the writing of the report; and in the decision to submit the paper for publication.

\section{Author Information}

\subsection{Contributions}

D.O. conceptualized, designed, and organized this study, analyzed the results, and wrote, reviewed,
and revised the paper. J.H. analyzed the results, and wrote, reviewed, and revised the paper.
R.K.P., J.C.P., and G.L.L. wrote, reviewed, and revised the paper. Y.P. reviewed and revised the
paper. Y.W. conceptualized, designed, and directed this study, wrote, reviewed, and revised the
paper.

\subsection{Corresponding author}

Correspondence to \href{mailto:yanshan.wang@pitt.edu}{Yanshan Wang}.


\section{Ethics declarations}
\subsection{Competing Interests}

Y.W. consults for Pfizer Inc. and has ownership/equity interests in BonafideNLP LLC. All other authors declare no competing interests.

\section{Data availability}
Not applicable.

\section{Code availability}
Not applicable.


\printbibliography


\end{document}